# Apparent Resonance Splitting in Self-Coupled Excitonic Systems

*Avishek Sarbajna[1], Qitong Li[2], Dorte Rubæk Danielsen[1], Skyler Peitso Selvin[2], Duc Hieu Nguyen[1], Manh-Ha Doan[1], Peter Bøggild[1], Mark L. Brongersma[2], Søren Raza[1,*]*

[1]Department of Physics, Technical University of Denmark, DK-2800 Kongens Lyngby, Denmark.

[2]Geballe Laboratory for Advanced Materials, Stanford University, Stanford, CA, USA.

[*]Corresponding author: sraz@dtu.dk

## Abstract

Thin films of high-refractive-index excitonic materials enable self-coupling by simultaneously supporting intrinsic excitonic transitions and optical resonances. These optical resonances take the form of Fabry–Perot resonances in thick films and absorption resonances in ultrathin films placed on metallic substrates. Here, we investigate whether these optical resonances lead to true exciton–photon hybridization. Using far-field reflectance and spectrally resolved photocurrent measurements, we study tungsten disulfide ($WS_2$) flakes on both metallic and dielectric substrates across a range of thicknesses. While reflectance spectra for ultrathin flakes exhibit resonance splitting between excitons and absorption resonances, our photocurrent measurements reveal only excitonic peaks, indicating that no polaritons are formed. In contrast, thicker flakes exhibit Fabry–Perot resonances that strongly couple to excitons, resulting in clear splitting in both reflectance and photocurrent spectra, and providing evidence of polariton formation. We further show that the polariton resonances can be tuned through the reflection phase at the $WS_2$–substrate interface by changing the substrate material. In addition to coupling with the strong A-exciton, we observe polariton formation involving the weaker B-exciton at shorter wavelengths, as well as higher-order hybridization where both excitons interact simultaneously with a single Fabry–Perot resonance. These findings clarify the distinction between apparent and true strong coupling in excitonic materials and demonstrate how reflection phase and flake thickness can be used to engineer light–matter interactions.

# Introduction

Self-coupled polaritonic systems—where a single material serves simultaneously as an optical cavity and the host of material resonances—offer an exciting platform for tuning light–matter interactions. In such systems, the formation of hybrid light–matter states[1-6] can significantly modify material properties, leading to enhanced device functionalities and applications ranging from photovoltaics[7] and polaritonic light-emitting diodes[8], to a broad suite of optoelectronic devices[9].

Excitonic and interband resonances are the two primary types of material resonances that have been explored in self-coupled polaritonic systems. The former have been used to couple to optical resonances in transition metal dichalcogenides (TMDs)[10-14], perovskites[15], and dye molecules[16-18], while the latter have been explored in metals, such as nickel[19-20] and in dielectric materials[21]. These diverse observations across material classes underscore the ubiquity of self-coupled polaritons[11].

Self-coupling in these systems is typically assessed through far-field techniques, such as reflectance spectroscopy. However, the broad spectral features of optical modes demand a careful interpretation of the coupling dynamics. In particular, for systems involving absorption resonances formed by placing a thin lossy high-index material on metallic substrates[22], earlier studies have interpreted the interaction with excitons as strong coupling[23-24]. Yet, the interaction between a broad optical resonance and a sharp excitonic transition can also produce Fano-type resonances with spectral features that resemble the resonance splitting of strong coupling, even though no polaritons are formed[25-27].

In this work, we investigate self-coupling between excitons and optical resonances in metal-supported tungsten disulfide ($WS_2$) flakes. $WS_2$ combines a high refractive index with strong excitonic absorption and hosts both A- and B-excitons that remain stable at room temperature[28-30]. By placing $WS_2$ on a metal substrate, Fabry–Perot (FP) resonances can be formed in thicker flakes, while ultrathin flakes (< 15 nm) exhibit absorption resonances due to nontrivial optical phase shifts at the $WS_2$-metal interface. To assess the self-coupling and whether resonance splitting corresponds to polariton formation, we perform far-field reflectance and spectrally resolved photocurrent measurements. While both FP and absorption resonances exhibit Rabi-like resonance splitting in reflectance measurements, only the FP resonance shows resonance splitting in photocurrent measurements consistent with exciton-polariton formation. Since photocurrent reflects absorption only within the active material ($WS_2$), we propose using the absorptance specifically within the active material as a more direct measure for assessing resonance splitting, rather than far-field reflectance measurements. Finally, we show that the reflection phase at the $WS_2$–substrate interface serves as a tuning parameter for the self-coupling, allowing control over the optical modes without altering the physical cavity length. These results clarify the distinction between apparent and true strong coupling and provide a strategy for designing compact, phase-engineered optoelectronic devices.

## Results

**Absorption and FP resonances in ultrathin WS$_2$ flakes**. Figure 1a shows an optical image of our sample which consists of an exfoliated WS$_2$ flake with thickness $t$ placed between two Au contacts on a SiO$_2$-on-Si substrate. The Au layer is 50 nm thick (Fig. 1b) and a 5 nm thick Cr layer is used to increase the adhesion between Au and SiO$_2$ (see Methods section). The Au contacts serve as both electrical contacts for the photocurrent measurements as well as a reflective substrate that modifies the optical resonances in the system. The transmission of light through the Au contacts is negligible, allowing the WS$_2$-on-gold to be treated as a separate system uninfluenced by the underlying SiO$_2$-on-Si substrate (see Fig. S1, Supplementary Note 1). This configuration allows us to assess the impact of both dielectric and metallic substrates on a WS$_2$ flake by locally illuminating different regions.

When the flake is significantly thinner than the optical wavelength inside WS$_2$ and placed on the reflective gold contact, the system functions as a lossy Gires–Tournois interferometer[31] (Fig. 1c). The finite optical conductivity of the gold contact along with the high refractive index and optical losses of WS$_2$ for wavelengths below the A-exciton (< 624 nm) leads to a destructive interference condition that gives rise to a dip in far-field reflectance and, consequently, a resonance in absorption[22][32] (see Supplementary Note 2). On the other hand, thicker WS$_2$ flakes can support FP resonances that can occur regardless of substrate material (Fig. 1d). However, the FP resonance wavelength depends on the substrate material due to the change in reflection phase at the WS$_2$-substrate interface.

To understand the coupling of excitons with these two different optical resonances, i.e., the absorption resonance in thinner flakes and the FP resonance in thicker flakes, we artificially nullify the oscillator strength of the A-exciton in WS$_2$ (see Methods, Fig. S2, Supplementary Note 3). This approach isolates the optical resonances, allowing their analysis without interference from exciton hybridization. Figure 1e-h shows the simulated absorptance of thin ($t$ = 12 nm) and thick ($t$ = 80 nm) WS$_2$ films on a gold substrate, when illuminated by a normally incident plane wave. With the excitonic contribution removed, we observe a single peak in absorptance due to the two different resonances (Fig. 1f, h). Upon interacting with the exciton, both of the resonances result in a resonance splitting (Fig. 1e,g) around the A-exciton at 624 nm (vertical dashed line).

Despite the strong similarity, it is important to evaluate the nature of the underlying light–matter interaction. In a strongly coupled system, the rate of energy exchange between an excitonic and optical resonance should exceed the combined rate of energy dissipation[33-34]. This condition is typically expressed by requiring that the Rabi splitting $E_\text{Rabi}$ exceeds the average full width at half maximum of the two resonances

$$E_{\text{Rabi}} > \frac{\Gamma_{\text{ex}} + \Gamma_{\text{cav}}}{2}, \tag{1}$$

where $\Gamma_{\text{ex}}$ and $\Gamma_{\text{cav}}$ denote the full width at half maximum of the excitonic and optical resonances, respectively. The second criterion for claiming strong coupling is that the coupling strength $g$ is

$$g > \frac{1}{4}|\Gamma_{\text{ex}} - \Gamma_{\text{cav}}| \tag{2}$$

To evaluate whether interaction between the excitons and optical resonances are in the strong coupling regime, we simulate the absorptance of WS$_2$ films on a gold substrate both including and excluding the contribution from the A-exciton (see Fig. 2). We observe that the absorption resonance exhibits a resonance splitting in the narrow thickness range of 10–14 nm, while the FP resonance enables exciton coupling over a broader thickness range (Fig. 2a). To determine the flake thicknesses of zero detuning, where the absorption resonance and FP resonance align spectrally with the exciton resonance, we analyzed the absorptance for varying thicknesses in the absence of the A-exciton (Fig. 2b). We find that the FP and absorption resonances spectrally match the A-exciton resonance at film thicknesses of $t = 81$ nm and $t = 14$ nm, respectively. At these film thicknesses, the Rabi splitting is evaluated for both optical resonances, and, by analyzing the corresponding absorptance spectra in the absence of the A exciton, we determine the FP and absorption resonance linewidths. The exciton linewidth is determined by considering a 5 nm thick flake where the optical response from the A-exciton is not influenced by either the absorptance or FP resonance. This enables us to retrieve the coupling strength. All extracted parameters to evaluate the strong coupling condition are presented in Table S2.

This analysis reveals that exciton-FP coupling meets both strong coupling criteria given by Eqs. (1-2), with the system exhibiting slightly more than one full Rabi oscillation[35] and thereby forming exciton-polaritons. On the other hand, the absorption resonance satisfies only the coupling strength criterion (Eq. (2)), while its Rabi splitting remains significantly smaller than the dissipation losses of the system (see Table S2). To further support this interpretation, we perform photocurrent measurements which confirm the same trend: exciton-polaritons emerge due to the FP resonance, whereas the absorption resonance, despite showing splitting in absorptance spectra, does not hybridize with the exciton.

**Far-field vs. photocurrent measurements.** We now investigate the optoelectronic properties of WS$_2$ flakes of varying thicknesses using photocurrent measurements. These measurements assess the system's external quantum efficiency (EQE), which quantifies how effectively incident photons are converted into charge carriers[36]. The EQE is defined as the ratio of the number of collected charge carriers (electrons or holes) to the number of incident photons at a given wavelength, and is experimentally calculated as

$$EQE(\lambda) = \frac{I_{\text{PC}}(\lambda)/e}{\Phi_{\text{photons}}(\lambda)}, \tag{3}$$

where $I_{\text{PC}}(\lambda)$ is the wavelength-dependent collected photocurrent, $e$ is the elementary charge and $\Phi_{\text{photons}}(\lambda)$ is the wavelength-dependent photon flux. This definition includes photons that may not interact with the active material due to reflection. The absolute EQE of our devices varies depending on the illumination wavelength and illumination area. Under a 1 V bias, thick $WS_2$ flakes on Au substrates (64 nm and 80 nm) achieve EQE values of 30–60 % across the 500–650 nm range, while thinner flakes (43 nm and 12 nm) yield EQE values below 10 % for the same spectral window. Since our primary interest lies in the relative spectral behavior and its comparison with far-field measurements, we normalize the EQE spectra to their maximum value and report the resulting quantity as the normalized EQE (NEQE) throughout the paper.

Figure 3a shows a schematic of our custom-designed photocurrent measurement setup. To enhance the photocurrent signal, we apply a bias voltage of 1 V during measurements. When probing flake segments on a $SiO_2$-on-Si substrate, the laser is focused approximately 2–5 µm away from the electrical contact (see Fig. S3 and Supplementary Note 4). A chopper is used to ensure that only the charge carriers generated by the laser illumination are measured. The photon flux is calculated from the power spectrum of the supercontinuum laser. The CCD is used to image the sample, and the photodetector collects the reflected laser light from the sample while performing photocurrent mapping (see Fig. S4, Supplementary Note 4).

For the experimental comparison between far-field derived absorptance and NEQE, we consider two $WS_2$ flakes thicknesses of $t = 12$ nm and $t = 80$ nm to investigate the impact of both absorption and FP resonances on exciton-polariton formation. The NEQE and absorptance measurements of these two flakes are presented in Fig. 3b-c. The absorptance of the system is determined from the experimental far-field reflectance $R_{\text{exp}}$ as $A_{\text{exp}} = 1 - R_{\text{exp}}$. For the thicker flake ($t = 80$ nm), experimental absorptance spectrum matches the NEQE spectrum (Fig. 3b), confirming our previous analysis about the formation of polaritons through FP-exciton coupling.

For the thin flake ($t = 12$ nm), NEQE measurements show no resonance splitting and we observe only peaks due to the A- and B-excitons (Fig. 3c). This is in stark contrast to the experimental absorptance spectrum taken from the far-field measurements, which shows resonance splitting at the A-exciton wavelength. This suggests that the interaction between exciton and absorption resonance does not generate polaritons. Further evidence for this claim is provided by photocurrent and absorptance measurements performed by illuminating the same flake on the $SiO_2$-on-Si substrate, i.e., between the electrical contacts (Fig. 3d). The absence of a reflective gold substrate removes the condition for destructive interference, eliminating the absorption resonance. Consequently, we do not observe resonance splitting in the absorptance spectrum (Fig. 3d). Comparison of the NEQE measurements taken by illuminating the $WS_2$ flake on and off the gold contact show only enhanced photocurrent from exciton excitation (Fig. 3c-d), with no evidence of polariton formation.

These results demonstrate that far-field derived absorptance spectra can misleadingly indicate resonance coupling, for example by showing apparent splitting even when no polaritons are formed. As a result, we do not find them to be a reliable metric for evaluating strong coupling. We therefore propose to use the absorptance only within the WS$_2$ flake $A_{WS_2}$ as a more direct measure of resonance coupling. This approach reflects the fact that photocurrent originates from absorption within the active material, making $A_{WS_2}$ a more relevant indicator than total absorptance. We simulate the flake absorptance by integrating the absorbed optical power across the WS$_2$ thickness[32] as

$$A_{WS_2} = \frac{1}{I_{inc}} \int_{z=0}^{z=t} \frac{\omega}{2} \text{Im}[\varepsilon] |E(z)|^2 \, dz, \qquad (4)$$

where $\omega$ is the angular frequency of the incident electromagnetic field, $|E(z)|^2$ is the internal electric field intensity at height $z$ measured from the flake-substrate interface, and $\text{Im}[\varepsilon]$ is the imaginary part of the permittivity of WS$_2$. The incident intensity is $I_{inc} = \frac{c\varepsilon_0}{2}|E_0|^2$, where $|E_0|^2$ is the field intensity of the normally-incident plane wave and $c$ is the speed of light in air. It is worth noting that this absorptance differs from the experimentally derived absorptance, which accounts for the entire system, including the substrate. In addition, this definition has shown good agreement in previous photocurrent studies[37-38].

The absorptance within the flake, $A_{WS_2}$, for different flake thickness and substrates are shown in Fig. 3b-d (dashed lines) along with the experimental results. For the thick flake ($t$ = 80 nm) on Au and the thin flake ($t$ = 12 nm) on SiO$_2$-on-Si, the resonance wavelengths of the simulations match almost exactly with the experimental outcomes (Fig. 3b,d). We observe that for the thin flake, the experimental absorptance increases at wavelengths above 650 nm (Fig. 3d). This increase is due to absorption in the silicon substrate and is therefore not captured in either the NEQE measurement or the $A_{WS_2}$ simulation.

For the thin flake on Au, the simulated absorptance within the WS$_2$ flake aligns significantly better with the NEQE measurements than the total device absorptance. Although the WS$_2$ absorptance does not show pronounced Rabi-like splitting, we observe a slight redshift of the absorptance maximum relative to the A-exciton wavelength, along with a broad peak between the A- and B-exciton regions. To understand these small differences between experiments and simulations, we examined several factors, such as the vertical position of absorption within the flake and the effect of incident angle and polarization, but found that none significantly affected the simulated spectra (see Fig. S5, Supplementary Note 5). While the resonance wavelengths in the NEQE and simulated WS$_2$ absorptance spectra are generally well aligned, the relative peak intensities do not always match. This discrepancy arises because absorptance simulations capture only the light absorption (i.e., charge generation) part of the photocurrent process, while the collection of those carriers is not accounted for. The latter depends on the flake–contact interfaces and flake conductivity. Variations in these parameters can affect the photocurrent response and thus alter the observed NEQE peak amplitudes. Nonetheless, our photocurrent measurements provide strong evidence that, unlike in far-field absorptance spectra, the presence of the absorption resonance does not lead to pronounced resonance splitting. This demonstrates that no exciton–polariton formation occurs in this regime.

**Impact of substrate and higher-order exciton-polaritons:** We now return to the self-coupling regime involving the FP resonance and excitons in thicker $WS_2$ flakes. By changing the substrate material, we demonstrate that the substrate-induced reflection phase of the optical resonance modifies the characteristics of self-coupled polaritons and enables coupling to the weaker B-exciton at shorter wavelengths. Varying the $WS_2$ film thickness further reveals the emergence of higher-order hybridization features. Together, these observations highlight the pronounced impact of polariton formation on the optoelectronic response of the material, beyond what is observed from excitons alone.

Figure 4a presents NEQE measurements of a thick flake ($t = 80$ nm), showing that changing the substrate from Au to $SiO_2$-on-Si causes the polariton peaks to redshift, as marked by the black arrows. This redshift is due to a change in the reflection phase in the FP resonance induced by the less reflective $SiO_2$-on-Si substrate, which makes the flake optically thicker compared to when it is on Au. The reduced amplitude of the long-wavelength polariton peak for flakes on $SiO_2$-on-Si is attributed to the lower extinction coefficient of $WS_2$ in this spectral region (see Fig. S2), a trend that is also reproduced in the simulated absorptance within $WS_2$ (Fig. 4b).

By decreasing the flake thickness to $t = 43$ nm, the FP resonance wavelength is blueshifted to the B-exciton wavelength, leading to the formation of self-coupled B-exciton polaritons (Fig. 4c). This is observed through a clear resonance splitting in both the NEQE measurements and the simulated $WS_2$ absorptance, demonstrating that even the weaker B-exciton is capable of hybridizing with the FP resonance. For an intermediate flake thickness ($t = 64$ nm) on Au substrate, the FP resonance is spectrally located between the A- and B-exciton resonances. In this case, both NEQE measurements and simulated absorptance reveal that the A- and B-exciton peaks are red- and blueshifted, respectively, relative to their uncoupled positions, and a third spectral feature appears near the bare FP resonance (Fig. 4d). A similar coupling scheme is theoretically examined in Ref. [39], where two excitonic states in a TMDC monolayer interact indirectly through a coupling to a single optical resonance, whose resonance energy lies between the two exciton energies. The indirect interaction gives rise to three mixed polariton branches, where the lower-energy B-like and higher-energy A-like branches are shifted downward and upward in energy, respectively, while a third, predominantly photonic branch remains near the uncoupled cavity energy. The spectral pattern we observe in Fig. 4d is consistent with this three-component hybridization, in which the FP resonance mediates the coupling between the A- and B-excitons. These NEQE measurements also match with the corresponding far-field experimental absorptance measurements (see Fig. S6, Supplementary Note 6).

**Self-coupled exciton-polariton formation:** In Figure 5, we summarize the formation of self-coupled exciton-polaritons across a wide range of $WS_2$ flake thicknesses. Resonant wavelengths from NEQE and experimental absorptance measurements are overlaid on absorptance simulations. Figure 5a-b show simulated and experimental results for flakes stacked on the gold and $SiO_2$-on-Si substrates, respectively. Depending on the optical response, we can divide the flakes into three different thickness regions. The exciton absorption enhancement region (region 1) has a very narrow thickness window (less than 15 nm) and occurs when the flakes are stacked on the gold substrate. Here, the structure supports an absorption resonance, which enhances absorption and shows a resonance splitting in experimental reflectance measurements. However, this splitting is not observed in NEQE measurements, indicating that no polaritons are generated. Region 2 is the uncoupled regime that covers a thickness range up to around 30 nm (for $SiO_2$-on-Si substrate) to 40 nm (gold substrate) where no interaction between optical resonances and excitons takes place. In region 3, which we identify as the exciton–polariton regime and which includes film thicknesses from 30 nm ($SiO_2$-on-Si) or 40 nm (Au) up to 100 nm, we observe hybridization between the FP resonances and both the A- and B-excitons. Resonance splitting is observed in both reflectance and photocurrent measurements. The three-component higher-order exciton-polariton are also present in the absorptance simulation and support our experimental outcome. Finally, we note that the flake thickness required to exhibit anti-crossing is different for metal and dielectric substrates because of the impact of the substrate-induced reflection phase pickup on the FP resonance wavelength.

## Conclusion

In this study, we investigated self-coupling between excitons and optical resonances supported by thin, lossy excitonic systems using $WS_2$ as the model material. Simulations and far-field measurements of device absorptance show that while coupling of absorption resonances to excitons in ultrathin $WS_2$ films can mimic Rabi-like splitting, they do not meet the criteria for strong coupling. This is confirmed in photocurrent measurements, which reveal the generation of excitons without polariton formation. In contrast, thicker flakes support Fabry–Perot resonances that strongly couple to the A-exciton and form exciton-polaritons, as evidenced by clear resonance splitting in both absorptance and photocurrent measurements. We further show that exciton-polaritons can be controlled through the reflection phase at the $WS_2$-substrate interface. By tuning either the substrate or $WS_2$ thickness, we demonstrate coupling to the weaker B-exciton as well as higher-order hybridization involving both the A- and B-excitons. These results clarify the conditions under which self-coupling occurs in $WS_2$ and establish reflection phase as an effective parameter in tuning light–matter coupling. These insights provide design rules for TMD-based optoelectronic devices, such as sensors, modulators and LEDs, and extend readily to other two-dimensional excitonic systems.

## Methods

**Sample fabrication:** The contacts are fabricated inside the cleanroom, starting with taking a highly p-doped 4-inch Si with 300 nm $SiO_2$ wafer. We spin coat the wafer with UV lithography polymer AZ 5214 E at 4000 rpm for 30 seconds to achieve 1.5 μm of thickness, followed by a soft bake at 90º for 120 s. Followingly we pattern the desired design onto the wafer using the UV lithography system MLA150 maskless aligner, with a 375 nm laser, from Heidelberg Instruments GmbH. The pattern is developed using AZ 726 MIF (2.38% TMAH in water) for 60 s and finally rinsed with DI water. We deposit Cr/Au, 5/50 nm respectively, using an e-beam evaporator system (Temescal). After the deposition, the wafer is soaked in acetone at 60º overnight for the lift-off process. Finally, the wafer is diced into smaller chips for use in this study. Thin flakes are mechanically exfoliated from $WS_2$ crystal from HQ Graphene using 3M Scotch Magic Tape 810 on a Si substrate and then transferred onto $SiO_2$-on-Si chips with contacts. For stacking we used a standard dry-transfer technique with a 10% polycarbonate (PC)/polydimethylsiloxane (PDMS) stamp and a transfer system from HQ Graphene (product code: HQ2D MOT). The stamp was heated to 110 °C to pick up the flakes from the Si substrate. For stacking on the $SiO_2$-on-Si chips, the temperature was raised to 200 °C, which melted the PC layer. Any residual PC was then removed by immersing the sample in chloroform for 10 minutes.

**Bright-field microscopy and reflectance spectroscopy:** The measurements are performed using a Nikon C2 confocal microscope and an unpolarized halogen light source for top illumination. For spectral accusation we used a 20x objective with 0.45 NA. The reflection optical images are taken by a Nikon DS-Fi1 camera. We used a confocal scanner and a 60 μm pinhole to spatially select the signal and then analyze it with a SpectraPro 2300i spectrometer (150 lines per millimeter) and Pixis Si charge-coupled device (−70 °C detector temperature). The reported reflection spectra are averaged by 10 frames, 0.05 s integration time each. The reflection spectra are normalized by the reflection spectra of a protected silver mirror (Thorlabs, PF10-03-P01) averaged and exposed similarly as the sample.

**Photocurrent measurements:** Photocurrent measurements are carried out utilizing a custom-built optoelectronic setup. A supercontinuum laser and an acousto-optic tunable filter (both from Fianium) are used to adjust the wavelength to have a monochromatic illumination (approximately 5 nm bandwidth). We use a mechanical chopper wheel operating at 400 Hz to modulate the laser light. The light is focused onto the sample using a 50x objective lens (Mitutoyo M Plan APO NIR, NA = 0.42, 20 mm working distance). The sample imaging is done using an imaging system comprising two 50:50 beam splitters mounted on flip mounts, a halogen lamp with a diffuser, and a CCD imaging camera equipped with a tube lens. A glass slide directs a small portion of the reflected laser light to a large-area Si photodiode (New Focus, model 2031), which is connected to a lock-in amplifier (Stanford Research SR810 DSP) for measuring the reflected signal. We use a three-axis piezo stage to precisely control the focused laser's spatial location on the sample. To extract the generated charge carriers, a DC bias is applied using a source meter (Keithley 2612) connected in series with a tunable current-to-voltage amplifier and the wire-bonded sample. The output voltage from the modulated amplifier is then fed to a second lock-in amplifier (Stanford Research SR810 DSP) to measure the photocurrent. We measure the power spectrum using a calibrated power meter (Thorlabs, PM-100USB) positioned at the sample's position. The spot size of the illumination arear is around 1 μm if not stated otherwise.

**Atomic force microscopy (AFM):** The thickness of the flakes is measured by a Dimension Icon-PT AFM from Bruker AXS in tapping mode.

**Simulation:** The absorptance simulations within WS$_2$ $A_{WS_2}$ are obtained using the finite-element method in COMSOL Multiphysics (Wave Optics module). The 2D model is constructed using periodic boundary conditions to reflect the in-plane translational symmetry and includes all the material layers in the device. The region above the WS$_2$ film is defined as air ($n$ = 1). Below the WS$_2$ film, we define a 50 nm Au layer, 5 nm Cr layer, a 300 nm SiO$_2$ layer and semi-infinite Si substrate. The refractive indices of Au, Cr, SiO$_2$ and Si are taken from ref. ([40], [41], [42], [43]), respectively. To mimic a semi-infinitely thick Si substrate we use an impedance boundary condition at the bottom boundary of the Si. When considering the dielectric substrate, we remove the metallic layers and simulate an air-WS$_2$-SiO$_2$-Si system. We use a port boundary to launch a normally-incident plane wave. The spectral step size is 1 nm. The anisotropic complex refractive index of WS$_2$ is taken from ref. ([44]). The optical response of the air/WS$_2$/Au system shown in Fig. 2 is simulated using an analytical three-layer model based on reflections and transmissions at air/WS$_2$ and WS$_2$/Au interfaces, including phase accumulation through the WS$_2$ film, while assuming normal incidence.

**Refractive index fit:** The experimentally measured permittivity of WS$_2$ in the spectral range 420-720 nm is fitted to a Lorentz model with four oscillator terms

$$\varepsilon(\omega) = \varepsilon_b + \sum_{j=1}^{4} \frac{f_j}{\omega_j^2 - \omega^2 - i\gamma_j \omega},$$

where $f_j$ represents the oscillator strength, $\omega_j$ is the resonance frequency, and $\gamma_j$ is the damping factor of the $j^{th}$ oscillator, The non-resonant contributions are accounted for through $\varepsilon_b$. The values of the fitted parameters along with a plot of the fit are shown in Table S1 and Figure S2, respectively. To eliminate the contribution from the A-exciton, we set the first oscillator strength to zero, i.e., $f_1 = 0$ (see Supplementary Note 3 and Figure S2).

All the colormaps used in the paper are from Ref. [45].

## Acknowledgements

**Funding**
A.S., D.R.D., P.B. and S.R. acknowledge support by the Independent Research Funding Denmark (1032-00496B). S.R. acknowledges support by Villum Fonden (VIL50376) and Novo Nordisk Foundation (NNF24OC0096142).

**Author contributions**
**Avishek Sarbajna:** Sample preparation and device fabrication, far-field optical measurements, photocurrent measurements, data analysis, theoretical framework, AFM measurements, writing original draft. **Qitong Li:** Photocurrent measurements. **Dorte Rubæk Danielsen:** Sample preparation and device fabrication. **Skyler Peitso Selvin:** Wire bonding the samples. **Duc Hieu Nguyen and Manh-Ha Doan:** Designing and fabricating Au contacts. **Peter Bøggild:** Co-supervision. **Mark L Brongersma:** Co-supervision, theoretical framework, revising manuscript. **Søren Raza:** Supervision, theoretical framework, writing original draft.

**Acknowledgement:**
We thank Prof. Simone Latini from Technical University of Denmark and Dr. Mohammad Taghinejad from Stanford University for fruitful discussions.

**Conflict of interest**
The authors declare no competing interests.

**Data and materials availability**
All the data and figures used in the paper, along with supplementary data, are available from the corresponding authors upon request.

**Figures**

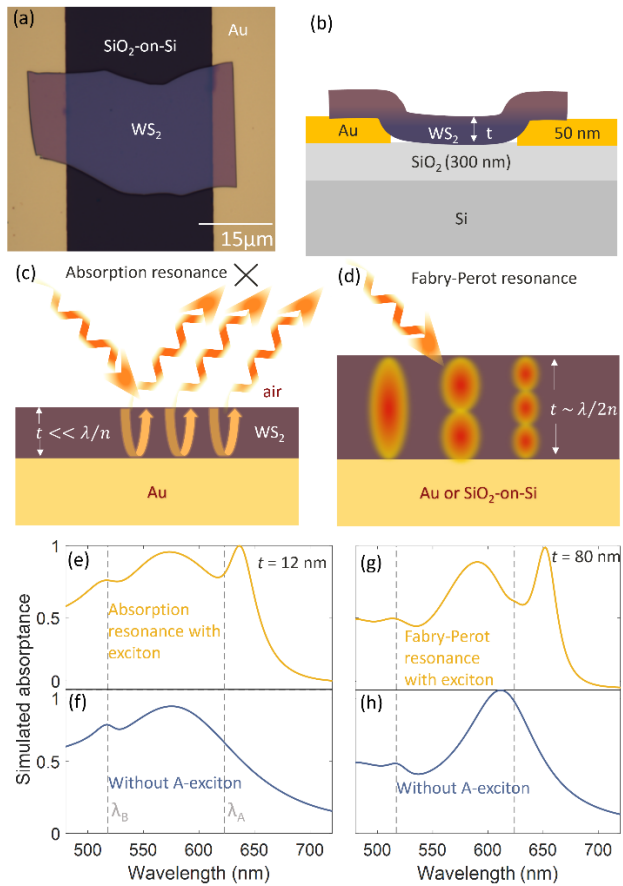

**Figure 1: The model system.** (a) Optical image (top view) and (b) a schematic (side view) of our sample. (c) Schematic of destructive interference that leads to an absorption resonance, when the flake is very thin and stacked on a metallic substrate. (d) Schematic of FP resonance supported by thicker flakes independent of substrate. (e-h) Simulated absorptance spectra of $WS_2$ flakes with thicknesses of $t = 12$ nm and $t = 80$ nm stacked on Au, respectively, shown with (e,g) and without (f,h) A-exciton contribution. In (f), the peak observed in the absence of the exciton is attributed to an absorption resonance, while in (h), the peak is due to the excitation of a FP resonance. The vertical dashed lines show the A- and B-exciton wavelengths at 624 nm and 517 nm, respectively.

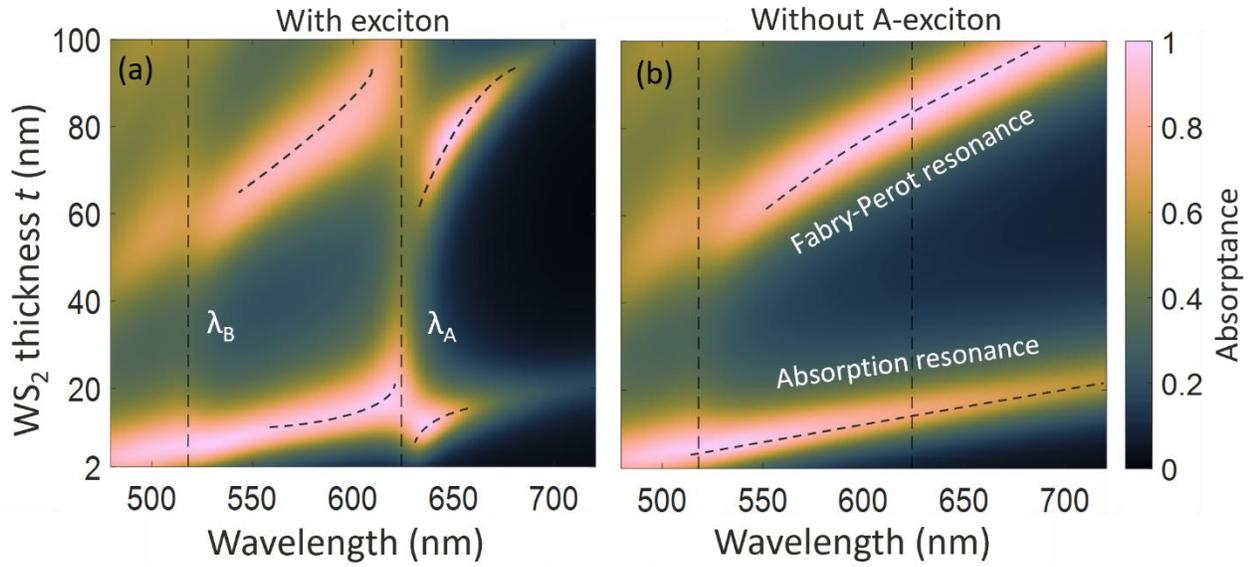

**Figure 2: Interaction between excitons and optical resonances.** (a-b) Simulated absorptance of WS$_2$ films on gold substrate both with and without the contribution from A-exciton, respectively. The black dashed curves are guides to the eye, and the vertical black dashed lines mark the A- and B-exciton wavelengths.

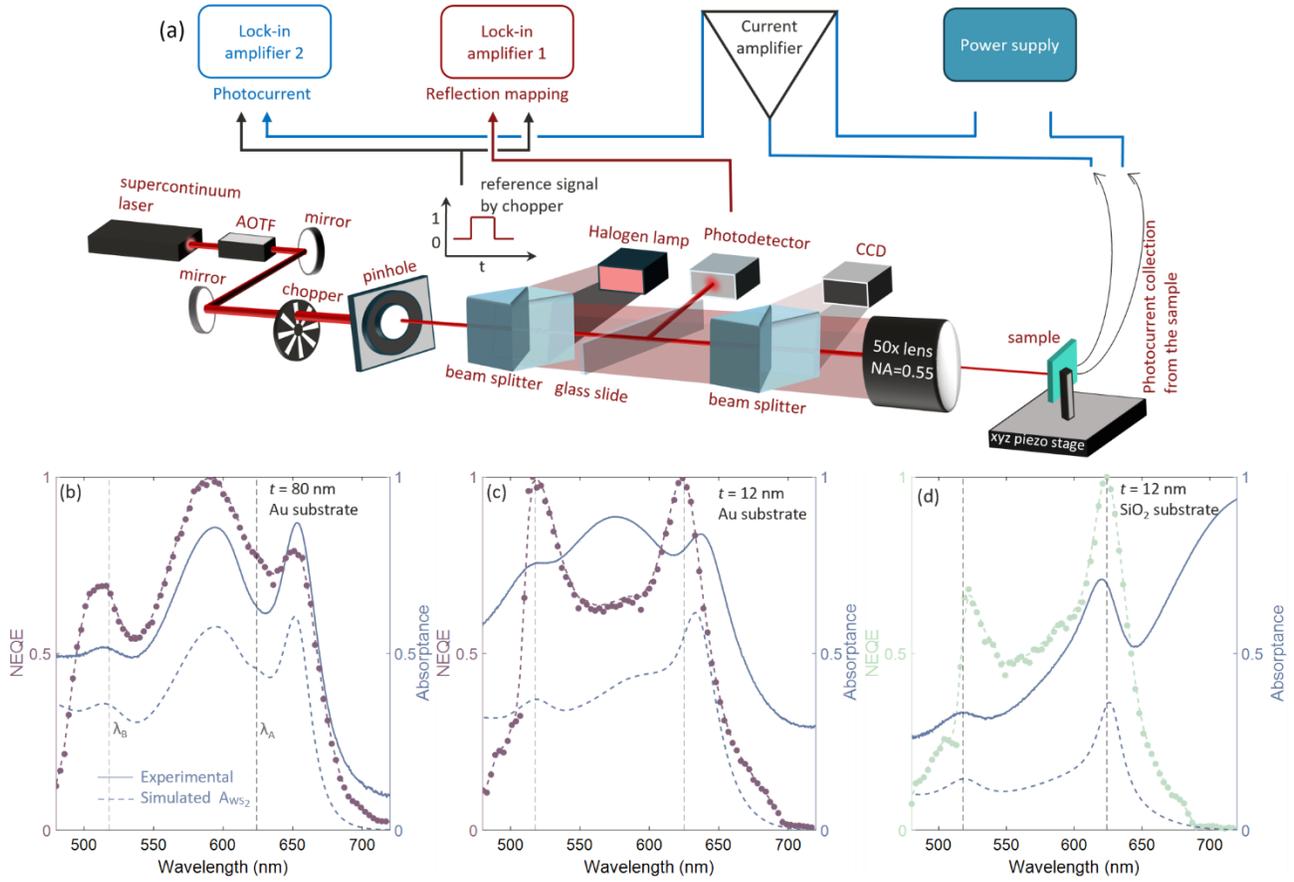

**Figure 3: Normalized EQE vs. absorptance.** (a) Schematic of photocurrent setup. The lock-in amplifiers receive the reference signal from the chopper and the signal from the photodetector, which are used to measure the photocurrent and reflected photons, respectively. The glass slide and the beam splitters are used to guide the laser to photodetectors (to collect reflected laser illumination from the sample) and the reflected halogen light to the CCD (for live view of the sample). (b-c) Experimental absorptance derived from reflectance measurements (blue), simulated absorptance within $WS_2$ given by Eq. (4) (dashed blue) and normalized EQE (purple) spectra for two $WS_2$ flake thicknesses of $t = 80$ nm and $t = 12$ nm stacked on Au substrate, respectively. (d) Similar measurements for $t = 12$ nm flake stacked on $SiO_2$-on-Si substrate. The vertical dashed lines show the spectral position of A- and B-excitons.

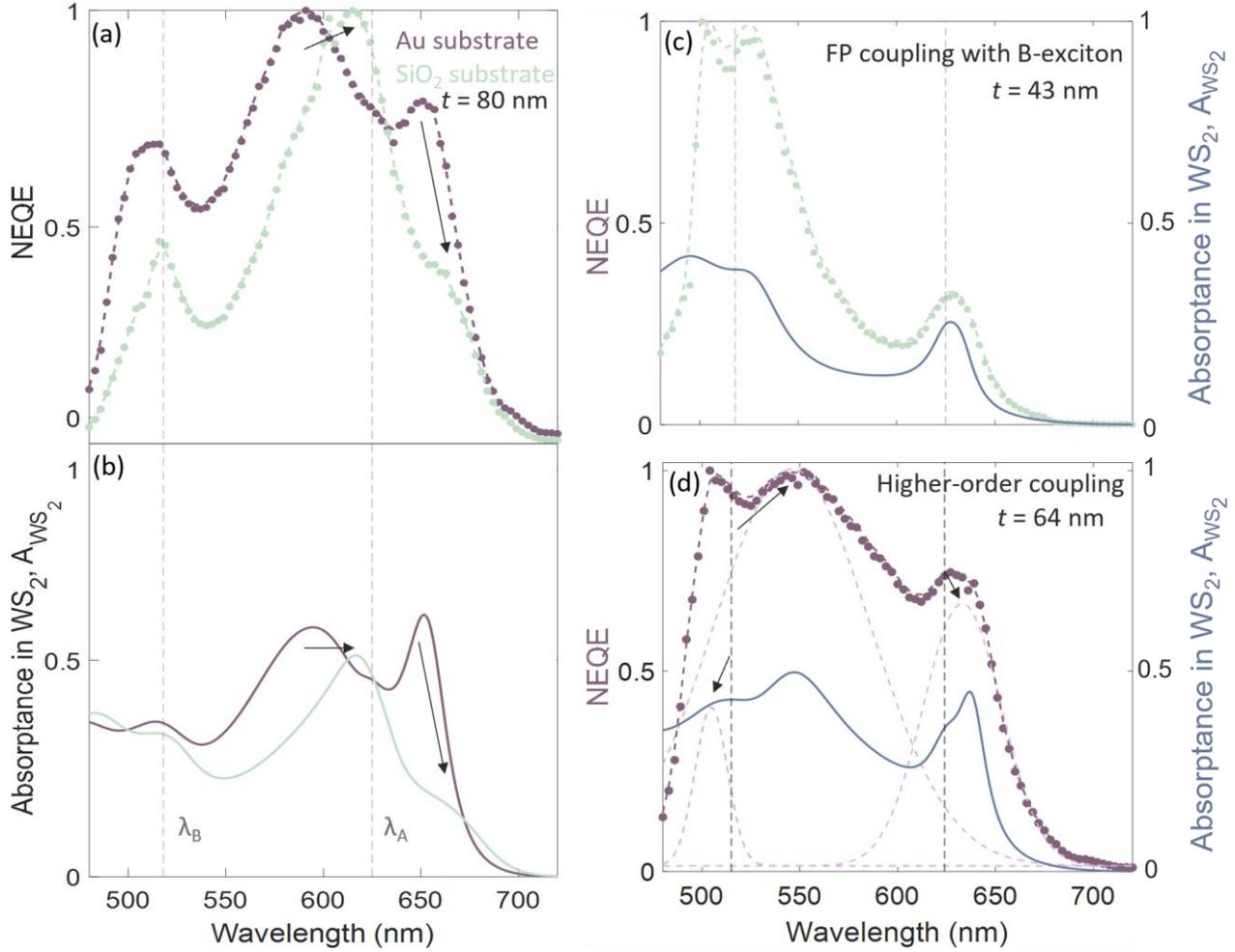

**Figure 4: Impact of substrate and higher-order exciton-polaritons.** (a-b) Comparison of NEQE measurements and simulated $A_{WS_2}$ for a WS$_2$ flake thickness of $t = 80$ nm on Au substrate (purple) and SiO$_2$-on-Si substrate (green). The shifts are shown by black arrows. (c) Coupling of Fabry–Perot resonance to the B-exciton in a WS$_2$ flake ($t = 43$ nm) on SiO$_2$-on-Si substrate. (d) Higher-order coupling between the Fabry–Perot resonance and both the A- and B-excitons in a WS$_2$ flake ($t = 64$ nm) on Au substrate. Lorentzian fits (light pink) are used to identify individual peak positions. The small peak at the A-exciton wavelength in the simulations are due to uncoupled excitons.

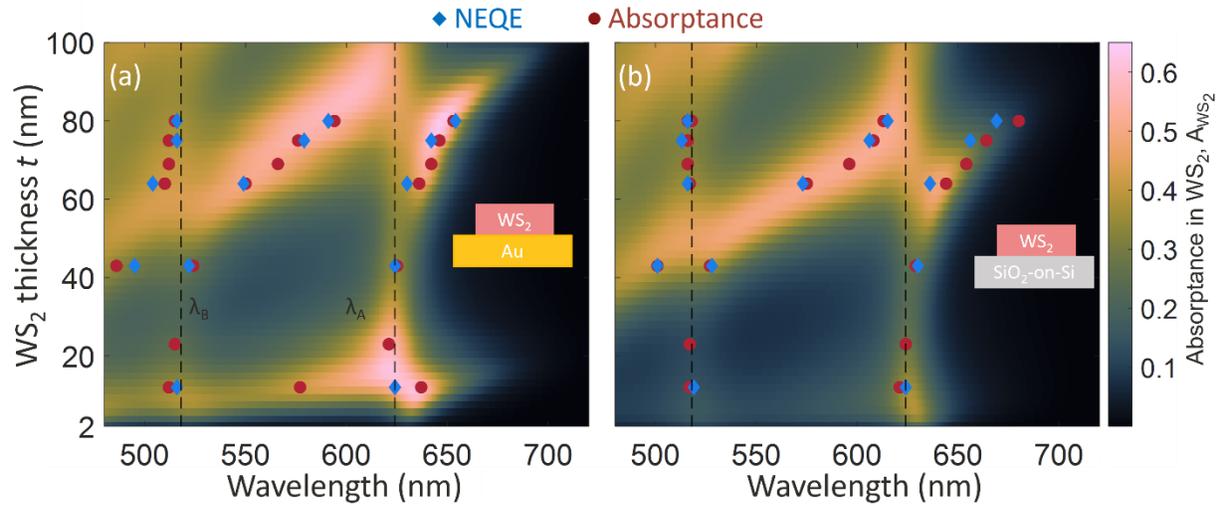

**Figure 5: Self-coupled exciton-polariton formation.** (a) Thickness-dependent simulations of $A_{WS_2}$ for flakes stacked on Au substrate and (b) on SiO$_2$-on-Si substrate. The blue symbols indicate the spectral positions of maxima from the NEQE measurements, and the red dots mark the maxima of the experimental absorptance peaks. The vertical dashed lines represent the spectral positions of the A- and B-exciton. Simulated absorptance is shown in the background for both cases. For some flake thicknesses, only reflection measurements were performed.

# Supporting Information for

# Apparent Resonance Splitting in Self-Coupled Excitonic Systems

*Avishek Sarbajna[1], Qitong Li[2], Dorte Rubæk Danielsen[1], Skyler Peitso Selvin[2], Duc Hieu Nguyen[1], Manh-Ha Doan[1], Peter Bøggild[1], Mark L. Brongersma[2], Søren Raza[1,*]*

[1]Department of Physics, Technical University of Denmark, DK-2800 Kongens Lyngby, Denmark.

[2]Geballe Laboratory for Advanced Materials, Stanford University, Stanford, CA, USA.

[*]Corresponding author: sraz@dtu.dk

**Supplementary Note 1: Gold contacts as a reflective surface**

The Au contacts in our system serve a dual purpose: they function as electrical contacts for photocurrent measurements and act as a reflective substrate that modifies the optical resonances within the structure. Due to the high reflectivity of gold, the optical interaction is primarily defined by the air-$WS_2$-gold system, with the underlying $SiO_2$-on-Si substrate having a negligible impact. Figure S1 presents a comparison of the simulated reflectance of $WS_2$ flakes placed on the experimental Au/Cr/$SiO_2$/Si substrate (50/5/300/∞ nm) with that of an idealized substrate consisting of an infinitely thick gold layer, showing excellent agreement between the two. This simplifies the analysis by decoupling the contributions from the $SiO_2$-on-Si substrate, allowing the $WS_2$-on-gold system to be treated independently.

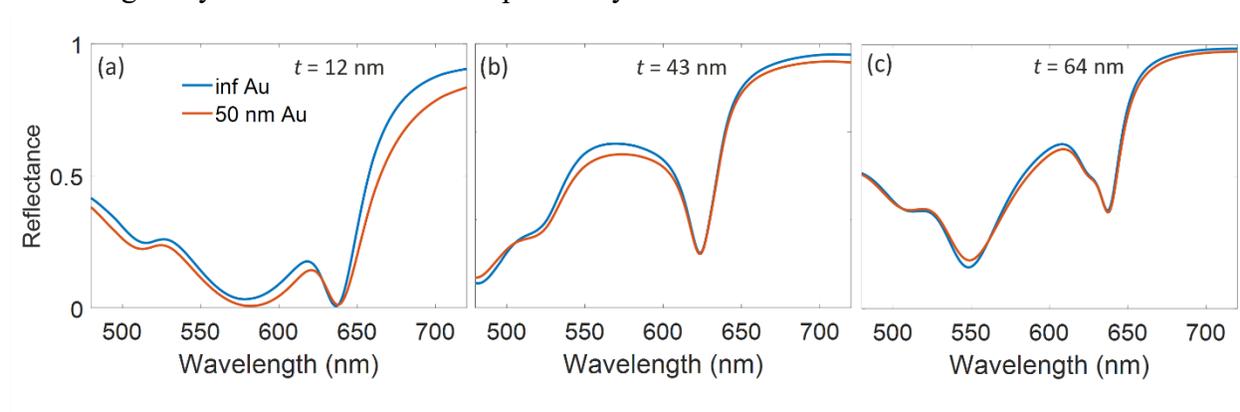

**Figure S1: Impact of Au substrate thickness:** Simulated reflectance spectra comparing $WS_2$ flakes of varying thicknesses (12 nm, 43 nm, and 64 nm) stacked on a Au/Cr/$SiO_2$/Si substrate (50/5/300/∞ nm) with an idealized substrate consisting of an infinitely thick gold layer.

## Supplementary Note 2: Absorption resonance

In this note, we derive the resonance condition for the absorption resonance and show that it arises in ultrathin films due to the finite optical conductivity of the metal mirror.

In general, a pronounced absorption resonance occurs when an ultrathin, lossy semiconductor of thickness $t$ is placed on a metal mirror with finite optical conductivity. The light reflected at the air-semiconductor interface interferes destructively with light that has undergone multiple propagations through the semiconductor layer and reflections from the metal mirror.

For a thin-film system sandwiched between a semi-infinite air superstrate and a semi-infinite metal substrate, the total reflection coefficient under normally-incident plane wave excitation is given by

$$r_{\text{tot}} = \frac{r_{\text{a-s}} + r_{\text{s-m}}\, e^{2i\beta}}{1 + r_{\text{a-s}} r_{\text{s-m}}\, e^{2i\beta}}, \tag{1}$$

where $r_{\text{a-s}} = (1 - \tilde{n}_s)/(1 + \tilde{n}_s)$ is the reflection coefficient at the air-semiconductor interface, $r_{\text{s-m}} = (\tilde{n}_s - \tilde{n}_m)/(\tilde{n}_s + \tilde{n}_m)$ is the reflection coefficient at the semiconductor-metal interface, and $\beta = (2\pi/\lambda)\tilde{n}_s t$ accounts for the propagation in the semiconductor layer. The complex refractive indices of the semiconductor and metal are $\tilde{n}_s = n_s + ik_s$ and $\tilde{n}_m = n_m + ik_m$, respectively.

To derive the condition for the absorption resonance, we consider a lossless semiconductor ($k_s = 0$) while retaining the finite optical conductivity of the metal. In this limit, Eq. (1) can be recast by expressing the interface reflection coefficients in polar form

$$r_{\text{tot}} = \frac{-|r_{\text{a-s}}| + |r_{\text{s-m}}|\, e^{i\left[2\left(\frac{2\pi}{\lambda}\right)n_s t + \phi_{\text{s-m}}\right]}}{1 - |r_{\text{a-s}}||r_{\text{s-m}}|\, e^{i\left[2\left(\frac{2\pi}{\lambda}\right)n_s t + \phi_{\text{s-m}}\right]}}, \tag{2}$$

where $\phi_{\text{s-m}} = \arg(r_{\text{s-m}})$ is the phase shift upon reflection at the semiconductor-metal interface. Setting $r_{\text{tot}} = 0$, we obtain the resonance condition from the phase constraint

$$2\left(\frac{2\pi}{\lambda}\right) n_s t + \phi_{\text{s-m}} = q(2\pi), \tag{3}$$

where $q$ is an integer. The first-order resonance ($q = 1$) corresponds to

$$t = \frac{\lambda}{4\pi n_s}(2\pi - \phi_{\text{s-m}}), \tag{4}$$

The result in Eq. (4) explains why the absorption resonance can occur in ultrathin semiconductor films. For a perfect electric conductor (i.e., infinite conductivity), the phase shift at the semiconductor-metal interface is $\phi_{\text{s-m}} = \pi$, which yields the Salisbury quarter-wavelength resonance condition, $t_S = \lambda/(4n_s)$. However, for a real metal with finite conductivity, such as Au at optical frequencies, the phase shift can exceed $\pi$ (i.e., $\phi_{\text{s-m}} > \pi$), allowing a resonance to occur at thicknesses significantly below $t_S$.

In the absence of loss in the semiconductor, the resonance leads only to modest absorption, as the metal provides the only loss channel. However, in the case of a strongly absorbing semiconductor (such as WS$_2$ in our study), the absorption resonance can lead to near-unity absorptance, with nearly all the light absorbed in the semiconductor layer [1-3].

**Supplementary Note 3: Fit of the WS$_2$ refractive index to the Lorentz oscillator model**

The dielectric function of WS$_2$ in the spectral range covering 420 to 720 nm can be fitted to a Lorentzian model with four oscillators (parameters are listed in Table 1).

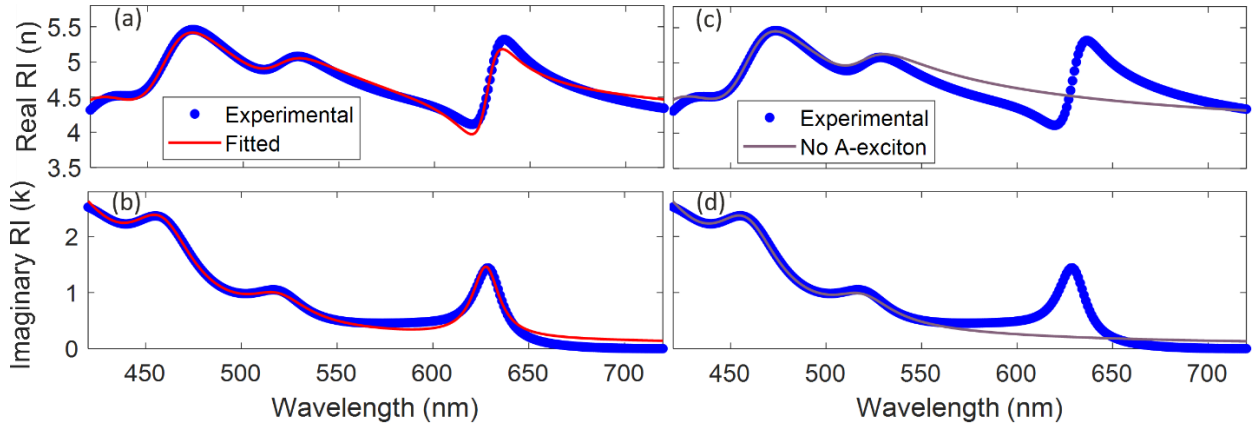

**Figure S2: Lorentz oscillator model for the refractive index of WS$_2$:** (a-b) Real and imaginary part of the experimentally measured and fitted refractive index of WS$_2$. (c-d) Refractive index after the artificial elimination of the A-exciton.

In Fig. S2a-b we show the real and the imaginary parts of the experimental and fitted refractive indices. The fit is based on the Lorentz model for the complex permittivity with four oscillator terms (see Methods in main manuscript). The first and second oscillator terms represent the A- and B-excitons, respectively. To eliminate the contribution of A-exciton, we set the first oscillator strength to zero, i.e., $f_1 = 0$ (see Fig. S2c-d).

**Table S1: Fitting parameters for Lorentz oscillator model**

| $\varepsilon_b$ | Oscillator no ($j$) | Oscillator strength ($f_j$) | Resonance frequency ($\omega_j$, eV) | Damping factor ($\gamma_j$, meV) |
|---|---|---|---|---|
| 12.04 | 1 | 1.18 | 1.97 | 50 |
| | 2 | 1.92 | 2.38 | 173 |
| | 3 | 8.98 | 2.69 | 230 |
| | 4 | 39.31 | 3.06 | 536 |

## Table S2: Parameters used to identify strong coupling

| Flake thickness (nm) | $E_{Rabi}$ (meV) | $E_{ex}$ (eV) | $E_{cav}$ (eV) | $\Gamma_{ex}$ (meV) | $\Gamma_{cav}$ (meV) | $\frac{1}{2}(\Gamma_{ex}+\Gamma_{cav})$ (meV) | $g$ (meV) | $\frac{1}{4}|\Gamma_{ex}-\Gamma_{cav}|$ (meV) |
|---|---|---|---|---|---|---|---|---|
| 14 | 130 | 1.97 | 2.07 | 49 | 750 | 399.5 | 180 | 175.5 |
| 81 | 190 | 1.97 | 2.01 | 49 | 290 | 169.5 | 110 | 60.5 |

## Supplementary Note 4: Photocurrent generation

We investigated how the photocurrent spectrum depends on the applied voltage. While increasing the voltage enhances the overall photocurrent signal through improved charge carrier collection, it does not alter the spectral shape. Figure S3a shows the photocurrent measured from a 64 nm thick flake stacked on Au under different applied biases. By normalizing the photocurrent spectrum at different applied bias, we observe negligible difference in relative amplitudes and that the peak wavelengths remain unchanged (Figure S3b).

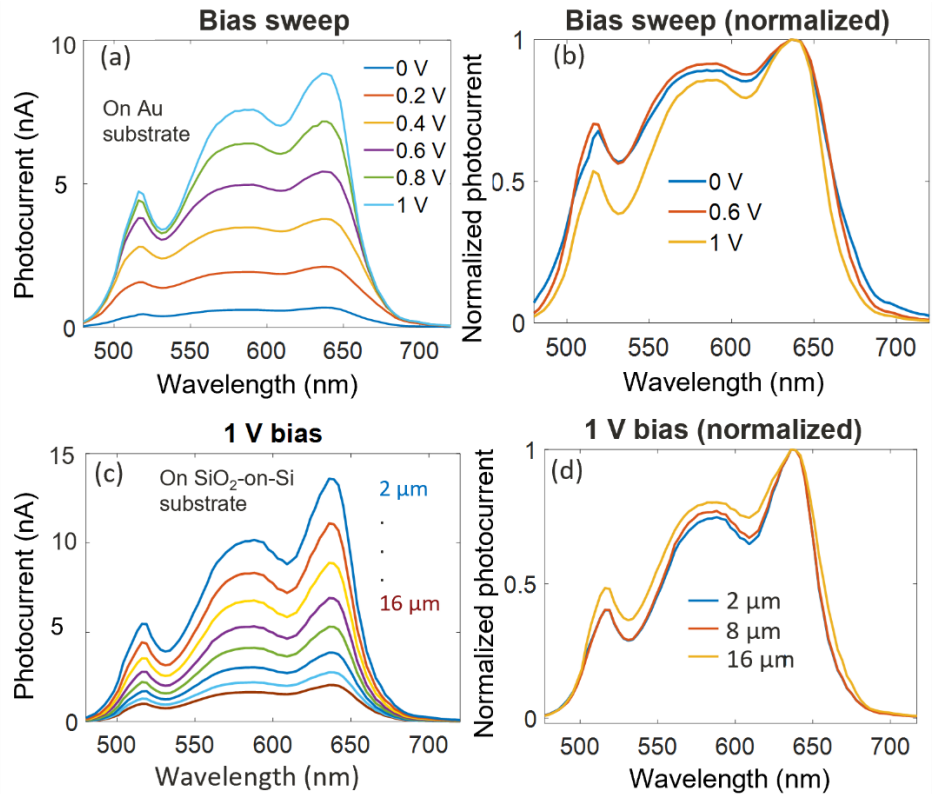

**Figure S3: Spectral shapes of photocurrent:** (a) Photocurrent measured under varying applied biases for Au substrate, showing higher photocurrent at higher biases. (b) Normalized photocurrent measured under varying applied biases for Au substrate, showing the relative change in amplitude at different applied biases. (c)

Photocurrent collected at different distances from the contact under 1 V bias, showing decreased photocurrent with increasing distance in both cases. (d) Normalized photocurrent collected at different distance from contact.

We also examined how the photocurrent spectrum depends on the distance between the laser spot and the gold contact for $WS_2$ flakes on a $SiO_2$-on-Si substrate. Placing the laser spot closer to the contact enhances charge carrier collection, leading to an increased photocurrent signal with only minor changes to the spectral shape. Figure S3c shows photocurrent collected from flake segment stacked on $SiO_2$-on-Si substrate at different distances from the contact (approximately starting from 2 to 16 µm in steps of 2 µm) under 1 V applied bias condition. Figure S3d shows normalized plots of the photocurrent measured at 2 µm, 8 µm and 16 µm from the Au contact showing almost no change in the relative amplitudes. Based on this analysis, we perform all our measurements at an applied bias of 1 V and close to the Au contact, when the flake is on $SiO_2$-on-Si substrate, to increase the photocurrent signal. Simultaneous maps of reflectance and photocurrent at 700 nm (Fig. S4a–b) verify that the photocurrent signal originates solely from the $WS_2$ flakes.

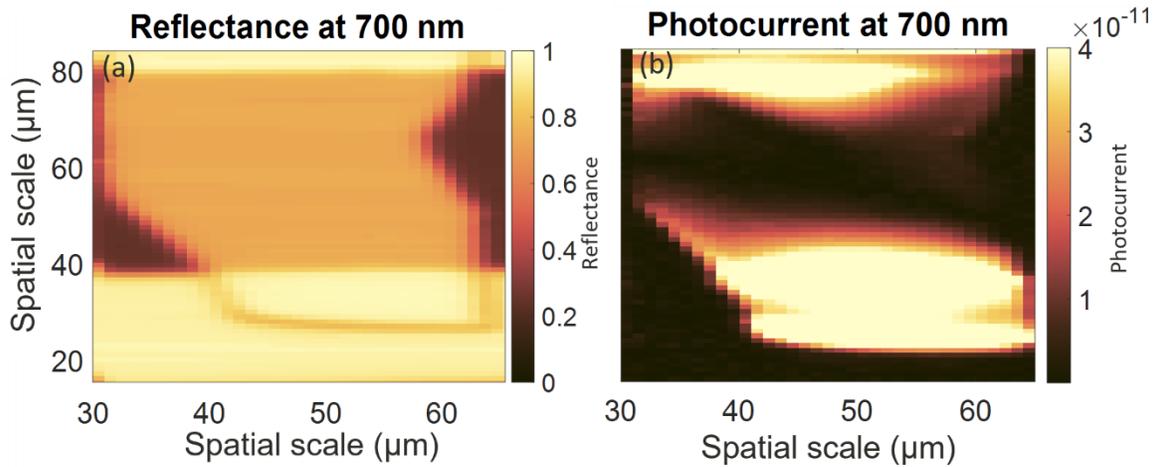

**Figure S4: Photocurrent mappings:** (a) Reflectance (b) photocurrent mapping of a 64 nm thick $WS_2$ flake under 700 nm laser illumination. Here we consider Au has 0.95 reflectance at 700 nm.

## Supplementary Note 5: 12 nm WS₂ flake on Au substrate

We considered several additional factors that might affect the simulated absorptance ($A_{WS_2}$) of the 12 nm thick flake and contribute to the observed redshift. We simulate $A_{WS_2}$ in different layer segments of the flake to analyze the contribution of different layers.

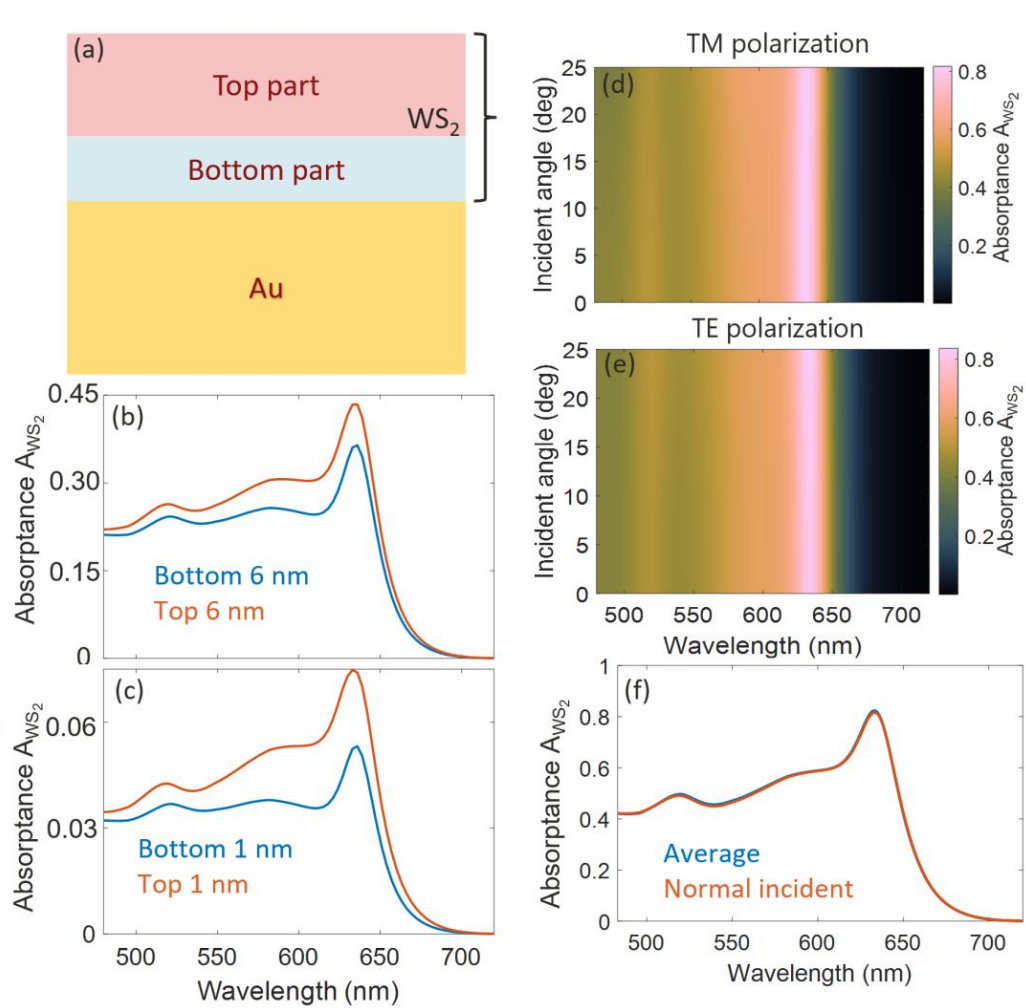

**Figure S5: Simulations for 12 nm flake on Au:** (a) Schematic of the simulation model dividing the 12 nm WS₂ flake into top and bottom parts. (b-c) $A_{WS_2}$ comparison between different sections of the flake. (d-e) $A_{WS_2}$ for TE and TM polarization of incident light as a function of incident angle. (f) Angle- and polarization-averaged $A_{WS_2}$ for 0–25° illumination, showing similar results to normal incidence.

Figure S5a shows a schematic of our simulation model, where a 12 nm thick WS₂ flake is divided into a top portion and a bottom part. We simulate $A_{WS_2}$ separately for each part while varying their individual thicknesses, keeping the total thickness fixed at 12 nm. Figure S5b compares the $A_{WS_2}$ calculated for the 6 nm thick top part versus the 6 nm thick bottom part. The $A_{WS_2}$ maxima in both cases occur at nearly the same spectral position, with a slight redshift relative to the A-exciton wavelength and a faint, broad peak between the A- and B-excitons. In addition, the top part consistently shows higher $A_{WS_2}$ than the bottom part. Figure S5(c) compares a 1 nm slice from the very top of the flake with one from the very bottom (in direct contact with the gold substrate); here, too, the spectral shapes are nearly identical. Thus, our simulations indicate that although the amplitude of $A_{WS_2}$ varies within the WS$_2$ layer, the spectral profile remains essentially unchanged.

Next, we simulate our experimental scenario where the illumination is limited by the numerical aperture (0.42) of the objective lens. Our goal is to determine whether the incident angle affects the spectral shape of the absorptance. Figure S5d–f present the simulated $A_{WS_2}$ of WS$_2$ integrated over illumination angles from 0° to 25°, for the two orthogonal polarization states: TE (transverse electric, where the electric-field vector lies along z axis, coming out of the plane in the 2D model) and TM (transverse magnetic, where the magnetic-field vector is along z axis). The results indicate that the angle-averaged simulated absorptance closely matches that obtained under normal incidence, with only a slight redshift of the absorption maximum.

**Supplementary Note 6: NEQE vs experimental absorptance**

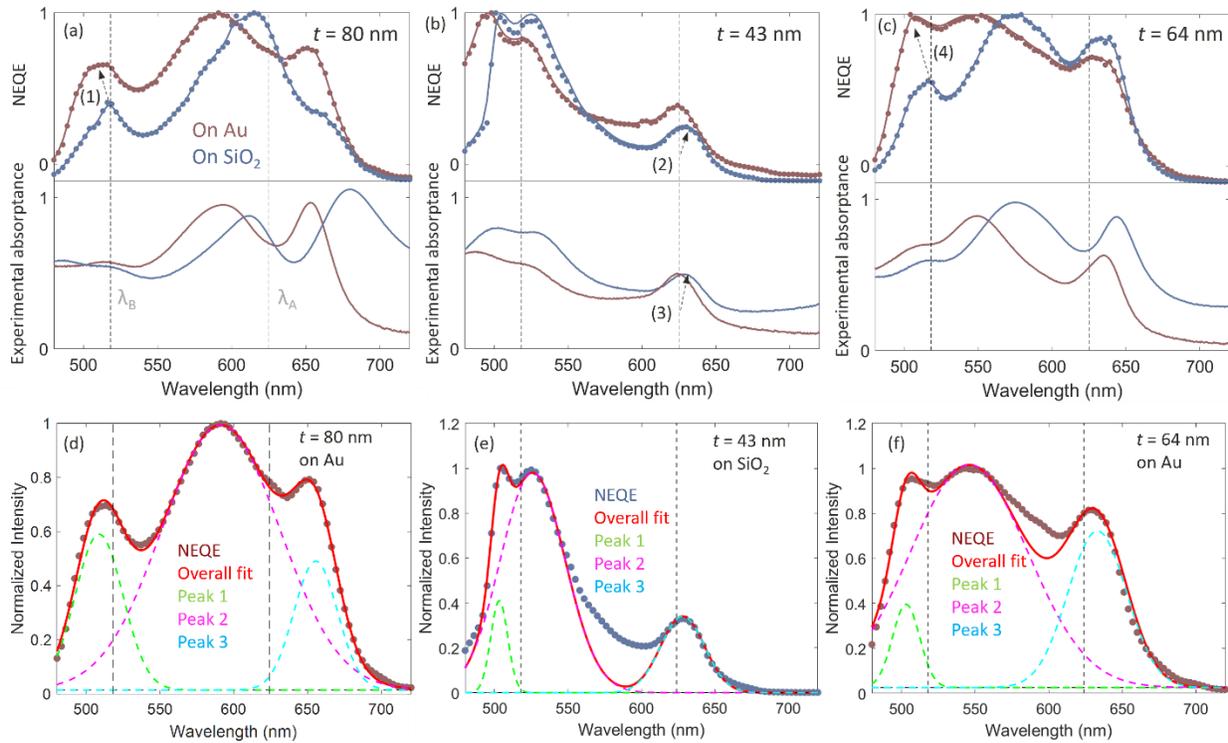

**Figure S6: Normalized EQE vs experimental absorptance:** (a-c) NEQE and experimental absorptance measurements for flake thickness $t$ = 80 nm, $t$ = 64 nm and $t$ = 43 nm, respectively. Plots in brown are from the flakes stacked on Au substrate and the plots in blue are from the same flakes on a SiO$_2$-on-Si substrate. Black arrows are used to indicate the spectral shifts of the maxima (minima) due to changing substrates. (d-f) Lorentzian peak fittings of NEQE spectra where higher-order coupling is observed. The vertical dashed lines show the spectral position of A- and B-exciton.

Figure S6a-c shows a comparison of the NEQE and experimental absorptance for the WS$_2$ flakes presented in Figure 4 of the main manuscript. For all three thicknesses, the NEQE exhibits a stronger B-exciton signature than the experimental absorptance. For the 80 nm flake on Au (Fig. S6a), we observe higher-order coupling resulting in a blueshift of the B-exciton in the NEQE measurements (indicated by black arrow (1)), although this feature is relatively weak in the absorptance spectrum. Conversely, for the flake on the SiO$_2$–on–Si substrate, the lower-wavelength polariton arising from coupling between the Fabry–Perot resonance and the A-exciton is more pronounced in the experimental absorptance than NEQE. The decreased NEQE signal at wavelengths above the A-exciton is due to the decreased material losses of WS$_2$ (see Figure S2). For the 43 nm flake (Fig. S6b), Fabry–Perot coupling to the B-exciton is observed for both substrates. Additionally, higher-order coupling is observed for the SiO$_2$–on–Si substrate (black arrows (2) and (3)). For the 64 nm thick flake, higher-order coupling is visible on the Au substrate but not on the SiO$_2$–on–Si substrate (Fig. S6c). To ensure that the observed spectral shifts are real and not resulting from peak overlapping, we fit the photocurrent spectrum with a sum of Lorentzian functions (Fig. S6d-f). The peak fitting of the NEQE spectra confirms that the wavelength shifts are indeed due to self-coupling between the FP resonance and the A- and B-excitons in WS$_2$.